\documentclass[11pt]{article}

\usepackage{amsfonts}
\usepackage{epsfig}
\usepackage{latexsym}
\usepackage{amsmath}
\usepackage{amssymb}
\usepackage{float}

\makeatletter
\@addtoreset{equation}{section}
\makeatother

\def\bequ{\begin{equation}}
\def\eequ{\end{equation}}
\def\barr{\begin{array}}
\def\earr{\end{array}}

\def\ben{\begin{equation}}
\def\een{\end{equation}}
\def\bena{\begin{eqnarray}}
\def\eena{\end{eqnarray}}

\def\b1{e^0}

\newcommand{\be}{\begin{equation}}
\newcommand{\ee}{\end{equation}}
\def\bea{\begin{eqnarray}}
\def\eea{\end{eqnarray}}

\newcommand{\beqc}{\begin{equation}\begin{gathered}}    
\newcommand{\eeqc}{\end{gathered}\end{equation}}

\def\nn{\nonumber}

\def\be{\begin{equation}}
\def\ee{\end{equation}}
\def\bea{\begin{eqnarray}}
\def\eea{\end{eqnarray}}
\def\beq{\begin{equation}}
\def\eeq{\end{equation}}

\def\ft#1#2{{\textstyle{\frac{\scriptstyle #1}{\scriptstyle #2} } }}
\def\fft#1#2{{\frac{#1}{#2}}}

\def\lesssim{\mathrel{\hbox{\rlap{\hbox{\lower4pt\hbox{$\sim$}}}\hbox{$<$}}}}
\def\gtrsim{\mathrel{\hbox{\rlap{\hbox{\lower4pt\hbox{$\sim$}}}\hbox{$>$}}}}

\newcommand{\hoch}[1]{$\, ^{#1}$}

\newcommand{\auth}{\Large\bf{A.H. Mujtaba\hoch{1} and C.N. Pope\hoch{1,2}}}

\begin{document}

\begin{flushright}
\hfill { 
MIFPA-12-42\ \ \
}\\
\end{flushright}

\begin{center}

{\Large{\bf The Hoop Conjecture
for Black Rings}}

\vspace{30pt}
\auth

\large

\vspace{20pt}{\hoch{1}\it George P. \& Cynthia W. Mitchell
Institute for\\ 
Fundamental Physics and Astronomy,\\ Texas A\& M University,
College Station, TX 77843-4242, USA}

\vspace{10pt}{\hoch{2}\it DAMTP, Centre for Mathematical Sciences,\\
 Cambridge University, Wilberforce Road, Cambridge CB3 OWA, UK}

\vspace{50pt}

\underline{ABSTRACT}
\end{center}

\vspace{15pt}

   A precise formulation of the hoop conjecture for four-dimensional
spacetimes proposes that the Birkhoff invariant $\beta$ for an apparent horizon
in a spacetime with mass $M$ should satisfy $\beta\le 4\pi M$.  The 
invariant $\beta$ is the least maximal length of any sweepout of the
2-sphere apparent horizon by circles.  An analogous
conjecture in five spacetime dimensions was recently formulated, 
asserting that the Birkhoff invariant $\beta$ for $S^1\times S^1$ sweepouts 
of the apparent horizon should satisfy $\beta\le \ft{16}{3}\pi M$.  Although
this hoop inequality was formulated for conventional five-dimensional
black holes with 3-sphere horizons, we show here that it is also obeyed
by a wide variety of black rings, where the horizon instead has $S^2\times S^1$
topology.

\pagebreak


\section{Introduction}

  Forty years ago, Thorne formulated the Hoop Conjecture, proposing that
in four dimensions an horizon forms when and only when a mass $M$ is 
compacted into a region whose circumference $C$ in every direction satisfies
the inequality $C\le 4 \pi M$ \cite{thorne}.  It is easily seen that the
Schwarzschild black hole saturates this bound, and thus the conjecture
would imply that other black holes of a given mass, such as those carrying 
charge or angular momentum, could not be ``larger'' than the Schwarzschild
black hole.  One of the problems
with testing the validity of the hoop conjecture is to make rigorous the
notion of the circumferences of the region in which the mass is confined.  In
an attempt to do this, Gibbons reformulated the hoop conjecture in terms
of a quantity called the Birkhoff invariant \cite{gib1}.  He considered a 
Cauchy surface $\Sigma$ containing an outermost marginally trapped surface
or apparent horizon $S$ with induced metric $g$, and constructed the 
Birkhoff invariant $\beta(g)$ for the pair $\{S,g\}$.  This is defined by
considering foliations of the topological 
2-sphere $S$, with its metric $g$,  as topological circles that
grow from a point (a ``north pole'') to a largest circumference 
(an ``equator'') and then shrink to a point again (a ``south pole'').  The
Birkhoff invariant $\beta(g)$ is then given by the smallest value amongst
all the equitorial circumferences that can be achieved by foliating $S$
in all possible ways.  The Gibbons reformulation of the hoop conjecture
in four dimensions then asserts that
\be
\beta(g)\le 4\pi M\,,\label{gibbons4}
\ee
with equality for the Schwarzschild black hole.  The validity of this
conjecture was demonstrated for various cases in \cite{gib1}, including
the charged and rotating Kerr-Newman black hole.

  Further tests of the inequality (\ref{gibbons4}) were performed in
\cite{cvegibpop},  where it was shown to be satisfied by more general
four-dimensional black holes, including those with a cosmological
constant, and various rotating (multi)charged black holes in ungauged and
gauged supergravities.  Generalisations of the hoop conjecture to higher
spacetime dimensions were also investigated in \cite{cvegibpop}.  It
becomes necessary then to find some appropriate definition of a 
higher-dimensional ``hoop,'' and then to give a precise definition of how
its size is to be calculated. Since a conventional black hole in
$D$ spacetime dimensions has an horizon with the topology of a $(D-2)$-sphere,
one natural generalisation of a hoop would be to consider foliating the
horizon with $(D-3)$-spheres, and define the analogous Birkhoff invariant
as the volume of the smallest equitorial $(D-3)$-sphere amongst all 
the possible such foliations.  There are, however, other natural possibilities.
For example, in $D=5$ spacetime dimensions, one could instead foliate 
a 3-sphere horizon with $S^1\times S^1$ Clifford torii, and consider the 
Birkhoff invariant defined as the smallest achievable 
``equitorial'' $S^1\times S^1$ area amongst all possible such foliations.  
These, and other possible formulations of hoop inequalities were explored
in \cite{cvegibpop}, and tests of the conjectures were performed for a
variety of rotating and charged black holes in higher dimensional 
gravities and supergravities.  All the $D$-dimensional 
black hole examples studied in \cite{cvegibpop} had the ``conventional''
$S^{D-2}$ horizon topology.

  In this paper, we shall study one of these hoop conjectures in 
five-dimensional
spacetime, but applied now to the black holes known as black rings,
which have horizon topology $S^2\times S^1$, rather than $S^3$ 
\cite{ring1}.  The formulation that is particularly appropriate
in this case uses the foliation by $S^1\times S^1$ torii mentioned above.
Now, however, instead of foliating 3-sphere horizons, the $S^1\times S^1$
torii are to be thought of as foliating the $S^2\times S^1$ horizons of the
black rings.  Thus one factor in the torus will be the circle of a 
foliation of $S^2$, while the other $S^1$ factor in the torus will be
the $S^1$ factor in the horizon.  Intriguingly, although the conjectured
inequality for $S^1\times S^1$ sweepouts in \cite{cvegibpop} was 
formulated only with five-dimensional black holes with $S^3$ horizon
topologies in mind, we shall see that the identical inequality is 
satisfied also by all the black-ring solutions that we have investigated.

\section{Hoop Inequality with $S^1\times S^1$ Sweepouts}

  It was conjectured in \cite{cvegibpop} that the Birkhoff invariant 
$\beta(g)$ for $S^1\times S^1$ sweepouts of an apparent horizon in a
five-dimensional spacetime satisfies the inequality
\be
\beta(g) \le \fft{16\pi}{3}\, M\,,\label{hoop5}
\ee
where $M$ is the mass.  The bound is exactly saturated by the five-dimensional 
Schwarzschild solution.  It was shown in \cite{cvegibpop} that the bound
is obeyed by the asymptotically flat rotating 3-charge black holes of ungauged
supergravity \cite{cveyou}, and by the asymptotically AdS rotating
charged black holes of five-dimensional minimal gauged supergravity 
\cite{cvchlupo}.  These, of course, all have horizons that are topologically
$S^3$.

Here, we shall examine the inequality (\ref{hoop5}) in the case of the original
uncharged black ring with a single rotation parameter \cite{ring1}, 
its charged generalisation \cite{elvang}, and the ring with two
rotation parameters that was obtained in \cite{pomsen}, together with
its charged generalisation \cite{hosk,galsch}.  In all cases, we shall
find that the inequality is obeyed.

\subsection{Single rotating uncharged black ring}

The metric for the single rotating uncharged black ring was first 
obtained in \cite{ring1}.   It was recast in a somewhat more convenient
form in \cite{emp2,emprea2}:
\beq
ds^2 = - \fft{F(y)}{F(x)} \left( dt + C R \, \fft{1 + y}{F(y)} d\psi \right)^2 
+ \fft{ R^2 }{ (x - y)^2 } F(x) \left[ - \fft{G(y)}{F(y)} d\psi^2 - 
\fft{ dy^2 }{G(y)} + \fft{ dx^2 }{G(x)} + \fft{G(x)}{F(x)} d\phi^2 \right]\,,
\label{uncharged}
\eeq
where $F(\zeta) = 1 + \lambda \zeta$, \hspace{4pt} 
$G(\zeta) = (1 - \zeta^2)(1 + \nu \zeta)$ \hspace{4pt} and 
\hspace{4pt} 
$\displaystyle C(\nu, \lambda) = \sqrt{ \lambda (\lambda - \nu) 
\fft{ 1 + \lambda }{ 1 - \lambda } }$. 
The apparent horizon is located at $y = -\nu^{-1}$. 
The coordinates are restricted to lie in the ranges 
$-1 \leq x \leq 1$ and $-\infty <y \le -1$. 
The requirement that $G(\zeta)$ have 
real roots implies that the parameters should be such that 
$0 < \nu \leq \lambda < 1$.

     To avoid conical singularities at $x = -1$ and at $y = -1$, one must 
respectively require that $\phi$ and $\psi$ be periodic with 
periods $\Delta\phi$ and
$\Delta\psi$ given by:
\beq
\Delta \phi = \Delta \psi = \fft{ 2 \pi \sqrt{ 1 - \lambda } }{ 1 - \nu }\,.
\eeq
A similar conical singularity at $x = 1$ is avoided by requiring that 
$\phi$ be periodic with $\displaystyle \Delta \phi = 
2 \pi \fft{ \sqrt{ 1 + \lambda } }{1 + \nu}$.  Reconciling the two 
periodicities 
therefore implies that the parameters $\nu$ and $\lambda$ must be related by
\beq
        \lambda = \fft{2 \nu}{1 + \nu^2}\,.\label{lambdasol}
\eeq

   The apparent horizon, which has the topology $S^2\times S^1$, is spanned
by the coordinates $x$ and $\phi$ on the $S^2$, and $\psi$ on the $S^1$.  
The $S^2$ is foliated by circles parameterised by $\phi$, with singular
orbits at $x=+1$ and $-1$, corresponding to the poles of the sphere.  Thus
we can obtain an upper bound on the value of the Birkhoff invariant by
maximising the $S^1\times S^1$ sweepout area
\beq
 \mathcal{A}(x) = \left. \Delta\phi \Delta\psi \sqrt{ g_{\phi\phi} \, 
  g_{\psi\psi} -g_{\phi\psi}^2 } \, \right|_{y = - \fft{1}{\nu}}
\eeq
as a function of $x$, for $-1\le x\le 1$. 

   Substituting in the components of the metric and noting the 
simplification $G(y = -1 / \nu) = 0$, we have
\beq \label{A0_x}
   \mathcal{A}(x) = \fft{ 4 \pi^2 R^2 (1 - \lambda) }{1 - \nu} 
\left[ \fft{ \nu \, C(\nu, \lambda)^2 }{ (\lambda - \nu) } 
\fft{ G(x) }{ (1 + \nu x)^2 } 
  \right]^{1/2}\,.
\eeq
The ADM mass is given by \cite{emp2,emprea2}
\beq \label{M_0}
        M = \fft{ 3 \pi R^2 }{4} \fft{ \lambda }{ (1 - \nu) }\,,
\eeq
and so verifying the hoop conjecture (\ref{hoop5}) amounts to showing that
\be
Z\equiv 1-  \left[ \fft{ 3 \mathcal{A}(x) }{ 16 \pi M } \right]^2 = 1- 
\fft{ \nu \, (1 - \lambda^2) }{ \lambda } 
\fft{ (1 - x^2) }{ (1 + \nu x) }\ge 0\label{Zinequality}
\ee
when $\lambda$ is given by (\ref{lambdasol}) and
\be
-1\le x\le 1\,,\quad \hbox{and}\quad 0<\nu\le 1\,.\label{ranges}
\ee
Since $\nu\le\lambda$, the inequality (\ref{Zinequality}) will be established
if we can show that
\be
P(x)\equiv (1-\nu^2) (x^2-1) +  1+ \nu  x \ge0\,.
\ee
Writing this as
\be
P(x) = \ft14 (3-\nu x) (x+\nu)^2 + \ft14 (1+\nu x)(x-\nu)^2\,,
\ee
and noting that $(3-\nu x)>0$ and $(1-\nu x)\ge 0$ when the conditions in
(\ref{ranges}) are
satisfied, we see that $P(x)$ is manifestly non-negative.  The 
apparent horizon of the uncharged
black ring (\ref{uncharged}) with a single rotation parameter therefore
obeys the hoop inequality (\ref{hoop5}).

  An alternative proof, which we shall present here to illustrate an
analogous one that we shall use later for the more complicated 
example of the charged black ring with two rotation 
parameters, is to take the expression for $Z$ in (\ref{Zinequality}), 
with $\lambda$ given by (\ref{lambdasol}), and parameterise $x$ and $\nu$
by
\be
x= -1 +\fft{2}{1+u}\,,\qquad \nu = \fft1{1+v}\,,\qquad
\ee
where $u\ge 0$ and $v\ge 0$.  It can then be seen that $Z$ is given by
\be
Z= \alpha (4+4u +10 v + 12 u v + 2 u^2 v+ 10 v^2 + 6 u v^2 + 4 u^2 v^2
+ 5 v^3 + 3 u^2 v^3 + v^4 + u^2 v^4)\,,\label{Z2}
\ee
where $\alpha$ is the non-negative quantity given by
\be
\alpha^{-1}=  (1+\nu^2)(1+\nu x) (1+u)^2 (1+v)^4\,.
\ee
Since the coefficient of every term in (\ref{Z2}) is positive, it follows
that $Z\ge0$ and the hoop bound is again verified. 

\subsection{Single rotating charged black ring}

   This solution was first obtained in \cite{elvang}.  For our purposes, it 
will be convenient to parameterise it somewhat differently, in a form that
reduces to (\ref{uncharged}) in the case that the charge parameter is set
to zero.  A simple way to obtain the solution in this form is in fact
to set to zero one of the rotation parameters in the more general 
doubly-rotating charged ring solution obtained in \cite{galsch}, and 
then to make
appropriate parameter and coordinate redefinitions to match those in 
(\ref{uncharged}).  This is achieved by making the replacements
\be
\nu=0\,,\quad \lambda\rightarrow\nu\,,\quad k\rightarrow 
 \fft{R}{\sqrt{2(1+\nu^2)}}\,,\quad \phi\rightarrow -\sqrt{1+\nu^2}\, \psi\,,
\quad \psi\rightarrow \sqrt{1+\nu^2}\, \phi
\ee
in the metric in \cite{galsch} (which appears in (\ref{2rotmet}) below).  
This gives the metric
\bea
        ds^2 &=& -D^{-2/3} \fft{F(y)}{F(x)} \left( dt + C 
\fft{(1 + y)}{F(y)} R \, c \, d\psi \right)^2 \\ 
&&+ D^{1/3} 
\fft{R^2}{(x - y)^2} F(x) \left[ - \fft{G(y)}{F(y)} d\psi^2 + 
\fft{G(x)}{F(x)} d\phi^2 + \fft{dx^2}{G(x)} - \fft{dy^2}{G(y)} \right]\,,\\
D &=& 1 + s^2 \fft{2 \nu (x - y)}{(1 + \nu^2) F(x)}\,,\label{Ddef}
\eea
where $c \equiv \cosh \delta$ and $s \equiv \sinh\delta$, with $\delta$
parameterising the charge.  The functions $F$ and $G$ are precisely those
defined under (\ref{uncharged}), and the constant $C$ is $C(\nu,\lambda)$
defined under (\ref{uncharged}), where in all cases $\lambda$ is
given by (\ref{lambdasol}).
The mass and conserved electric charge for this solution are given by
\be
M= (1+\ft23 s^2)\, M_0\,,\qquad Q= 
\fft{\pi R^2\nu s c}{(1-\nu)(1+\nu^2)}\,,
\ee
where $M_0$ is the mass of the uncharged black ring, given in (\ref{M_0}).

  The apparent horizon is again located at $y=-\nu^{-1}$, and the area of the 
$S^1\times S^1$ sweepout on an $x=\hbox{constant}$ latitude of the $S^2$
is given by
\be
 \mathcal{A}(x) = \fft{c}{D^{1/6}} \mathcal{A}_0(x)\,,
\ee
where $\mathcal{A}_0(x)$ is the area of the sweepout in the uncharged
case ($\delta \rightarrow 0$) and therefore is given by the 
expression (\ref{A0_x}).  Thus
\be
 \fft{ 3 \mathcal{A}(x) }{ 16 \pi M } = \fft{c}{ D^{1/6} 
  (1 + \fft{2}{3} s^2 ) } \left[ \fft{ 3 \mathcal{A}_0(x) }{ 16 \pi M_0 } 
\right]\,,
\ee
and since the hoop inequality (\ref{hoop5}) 
has already been verified for the uncharged case, we need only show that
\be
 \fft{c}{ D^{1/6} (1 + \ft{2}{3} s^2) } \leq 1
\eeq
for $-1 \leq x \leq 1$, $0 < \nu \leq 1$ and all $\delta$ in order to
verify it for the charged case too.
 
  Since $x\ge -1$ and $y\le -\nu^{-1}<-1$, it follows that $(x-y)>0$.  Also,
$F(x) = 1 + 2\nu x/(1+\nu^2)$ and so with $0<\nu\le 1$ it follows that
$F(x)\ge 0$.  Hence, from (\ref{Ddef}), we see that $D>1$, and so it remains 
only to show that $c/(1+\ft23 s^2)\le1$.  This is clear, since
\be
\fft{c}{1+\ft23 s^2}= \fft{3c}{1+2c^2}= 1- \fft{(c-1)(2c-1)}{1+2c^2}\,,
\ee
and $c=\cosh\delta\ge1$.  Thus the hoop inequality (\ref{hoop5}) is 
satisfied by the single rotating charged black ring.

\subsection{Doubly rotating charged black ring}

   The solution for an uncharged doubly rotating black ring  was obtained 
in \cite{pomsen}.   This was generalised \cite{hosk} to a 2-charged doubly 
rotating black ring, and further in \cite{galsch} to the more general
3-charge doubly-rotating black ring solution in the ${\cal N}=2$ STU 
supergravity theory.  It was shown in \cite{galsch} that in order to
obtain a solution with no conical singularities at the poles of the
$S^2$ surfaces, two of the three charges must be set to zero. 
The metric is then given by
\bea
        ds^2 &=&  \, - \, D^{-2/3} \fft{ H(y,x) }{ H(x,y) } 
 (dt \, + \, c \, \Omega)^2 \, + \, D^{1/3} \bigg( - \, 
 \fft{ F(x,y) }{ H(y,x) } d\phi^2 \, - \, 2 \fft{ J(x,y) }{ H(y,x) } 
  d\phi d\psi \, \nn\\
&&+ \, \fft{ F(y,x) }{ H(y,x) } d\psi^2 \, + 
  \, \fft{ 2k^2 H(x,y) }{ (x \, - \, y)^2 (1 \, - \, \nu)^2} 
 \left[ \fft{dx^2}{G(x)} \, - \, \fft{dy^2}{G(y)} \right] \bigg)
\label{2rotmet}
\eea
where
\bea
        G(x) &=& (1 - x^2)(1 + \lambda x + \nu x^2) \,,\nn\\
        H(x,y) &=& 1 + \lambda^2 - \nu^2 + 2 \lambda \nu (1 - x^2) y + 
    2 x \lambda (1 - y^2 \nu^2) + x^2 y^2 \nu (1 - \lambda^2 - \nu^2)\,,\nn\\
  J(x,y) &=& \fft{ 2k^2 (1 - x^2) (1 - y^2) \lambda 
\sqrt{\nu} }{(x - y) (1 - \nu)^2} [1 + \lambda^2 - \nu^2 
  + 2(x + y)\lambda\nu - x y \nu( 1 - \lambda^2 - \nu^2)] \,,\nn\\
        F(x,y) &=& \fft{2k^2}{(x - y)^2 (1 - \nu)^2} 
\bigg\{ G(x)(1 - y^2) \bigg([(1 - \nu)^2 - \lambda^2](1 + \nu) + 
 y \lambda (1 - \lambda^2 + 2 \nu - 3 \nu^2) \bigg) \nn\\
 &&+ \, G(y) \bigg[ 2\lambda^2 + x \lambda [(1 - \nu)^2 + \lambda^2 ] + x^2 [ (1 - \nu)^2 - \lambda^2 ] (1 + \nu) \nn\\
 &&\hspace{40pt}+ \, x^3 \lambda ( 1 - \lambda^2 - 3 \nu^2 + 2 \nu^3 ) - x^4 (1 - \nu) \nu (-1 + \lambda^2 + \nu^2) 
   \bigg] \bigg\} \,,
\eea
together with 
\bea
        \Omega &=& - \fft{ 2k\lambda \sqrt{ (1 + \nu)^2 - \lambda^2 } }{ H(y,x) } 
\Big[\fft{1 + y}{1 - \lambda + \nu} 
 [1 + \lambda - \nu 
 + x^2 y \nu (1 - \lambda -\nu) + 2 \nu x (1 - y) ] 
 d\phi \nn\\
&& \qquad\qquad\qquad\qquad\qquad
    +(1 - x^2) y \sqrt{\nu} d\psi  \Big] \,,\nn\\
D &=& 1 + \fft{2 \lambda s^2\, (1 - \nu) (x - y) (1 - \nu x y)}{ H(x,y) }
\,,
\eea
where $-1 \leq x \leq 1$, $y \leq -1$ and $\phi$ and $\psi$ each have period
$2\pi$. 
The parameters are restricted by the requirements
$0 \leq \nu < 1$ and $2 \sqrt{\nu} \leq \lambda < 1 + \nu$.
The apparent horizon is located at
\be
        y_h = \fft{ - \lambda + \sqrt{ \lambda^2 - 4 \nu }}{ 2 \nu }\,,
\ee
and the ADM mass is given by
\be
        M = \fft{(3+2s^2) k^2 \pi \lambda }{ (1 - \lambda + \nu) }\,.
\ee

  As in the previous examples, we consider the family of $S^1\times S^1$
sweepouts of the horizon that are parameterised by the coordinate $x$. 
The area of the sweepout is given by
\be
        \mathcal{A}(x) = \left. (2 \pi)^2 \sqrt{ g_{\phi\phi} \, 
  g_{\psi\psi} -g_{\phi\psi}^2 } \, \right|_{y = y_h}\,.
\ee
Writing
\be
 \left[ \fft{ 3 \, \mathcal{A}(x) }{ 16 \pi M } \right]^2 
  \equiv D^{-1/3} \, Y\,, \label{2rothoop}
\ee
the verification of the hoop conjecture (\ref{hoop5}) amounts to showing
that $D^{-1/3} \, Y \le 1$.

  The algebra in this example is rather too complicated to present.  However,
the verification of the hoop conjecture may be demonstrated as follows. The
strategy is to reparameterise the constants $\nu$, $\lambda$ and $\delta$, 
and the latitude coordinate $x$ on the 2-spheres, so that each parameter
ranges, unrestricted, over the non-negative range 0 to $\infty$. We do this
by first defining
\be
\nu = \tanh^2\beta\,,\qquad \lambda= 2\tanh\beta\, \cosh\alpha\,.
\ee
The parameter $\beta$  lies in the range $0\le\beta\le\infty$,
while $\alpha$ should range over $0\le\alpha\le \alpha_{\rm max}$, where
\be
e^{\alpha_{\rm max}} = \coth\beta\,.
\ee
Note that the horizon will be located at $y_h=- e^{-\alpha}\, \coth\beta$.
Finally, we introduce new parameters $(u,v,w,z)$, each lying in the
range from 0 to $\infty$, such that
\bea
e^\beta &=& 1+u\,,\quad e^\alpha = 1 + \fft{2}{(1+v)[(1+u)^2-1]} \,,\quad
x= -1 +\fft{2}{1+w}\,,\quad
e^{\delta}= 1+z\,.
\eea
This gives a complete covering of the parameter space, and the $x$ coordinate
range on the 2-sphere, in terms of independent and unrestricted non-negative
quantities.  (We can, without loss of generality, assume that 
$\delta\ge0$ since the metric is insensitive to the sign of the 
electric charge.) 

We then find that $1-Y$ is a rational function of $u$, $v$, $w$ and $z$, in 
which every term in the numerator and denominator multinomials has a 
positive coefficient.  (For example, the numerator is a multinomial with
3350 terms, all having positive coefficients.)  This establishes that $Y\le 1$.
Likewise, we find that $D-1$ is a rational function with all positive 
coefficients, and so $D\ge 1$.  Therefore, from (\ref{2rothoop}), we see that
the hoop conjecture is verified for the single-charged doubly rotating
black ring.

\section{Conclusions}

   A formulation of a hoop conjecture for apparent horizons in five-dimensional
spacetimes (\ref{hoop5}) 
was proposed in \cite{cvegibpop}, based on a definition of a
``hoop'' in terms of a Birkhoff invariant for a 
least maximal sweepout by $S^1\times S^1$ foliations
of the three-dimensional horizon.  This extended an earlier reformulation
of Thorne's original \cite{thorne} four-dimensional hoop conjecture by
Gibbons, in which the hoop was characterised by the Birkhoff invariant for
$S^1$ foliations of the $S^2$ horizon \cite{gib1}.  The conjecture in
five spacetime dimensions was showen to be valid for various known black
hole solutions in \cite{cvegibpop}, but only those with $S^3$ horizon 
topology were investigated there.  In the present paper, we have examined the
conjecture also for black ring metrics, where the horizon topology is
instead $S^2\times S^1$.  Intriguingly, we find that the identical inequality
is obeyed in these examples also.

   In closing, we note that although the Gibbons reformulation \cite{gib1} 
of the
original Thorne conjecture \cite{thorne} has the great merit of replacing the
rather heuristic and loosely-defined notion of bounding the size of an 
apparent horizon ``in all directions'' by the precise notion of a Birkhoff
invariant for the horizon, it does come at the price of considerably
weakening the original concept of a hoop bound.  Thus, instead of asserting
that the black hole could be ``passed through the hoop'' with any orientation,
it asserts that there exists some orientation for which it would pass
through the hoop.  If, for example, in four dimensions 
the horizon had the shape of a
prolate ellipsoid, then the Birkhoff invariant would be given by the more
slender circumference around the equator of the ellipsoid, leaving open the
possibility that the dimension of a circumference passing through the
poles might exceed the conjectured hoop bound.  It would be very interesting
to investigate whether well-defined conjectures that encompassed 
the possibility of more powerful bounds in the spirit of the original hoop
conjecture might be formulated.

\section*{Acknowledgements}

C.N.P. is supported in part by DOE grant DE-FG03-95ER40917.


\begin{thebibliography}{99}

\bibitem{thorne} K.S. Thorne, 
{\it Nonspherical gravitational collapse: A short review} in 
{\it Magic without Magic} ed. J. Klauder (San Francisco: Freeman) (1972).  

\bibitem{gib1} G.W. Gibbons, 
{\it Birkhoff's invariant and Thorne's hoop conjecture},
arXiv:0903.1580 [gr-qc].

\bibitem{cvegibpop} M. Cveti\v c, G.W. Gibbons and C.N. Pope,
{\it More about Birkhoff's invariant and Thorne's hoop conjecture for
horizons},
Class.\ Quant.\ Grav.\  {\bf 28}, 195001 (2011),
arXiv:1104.4504 [hep-th].

\bibitem{ring1} R. Emparan and H.S. Reall, {\it A rotating black ring in 
five dimensions}, Phys.\ Rev.\ Lett.\ {\bf 88}, 10 (2002), hep-th/0110260.

\bibitem{cveyou} M. Cveti\v c and D. Youm,
{\it General rotating five-dimensional black holes of toroidally 
compactified heterotic string},
Nucl.\ Phys.\ {\bf B476}, 118 (1996),  hep-th/9603100.

\bibitem{cvchlupo} Z.-W. Chong, M. Cveti\v c, H. L\"u and C.N. Pope,
{\it General non-extremal rotating black holes in minimal five-dimensional 
gauged supergravity},
Phys.\ Rev.\ Lett.\  {\bf 95}, 161301 (2005), hep-th/0506029.

\bibitem{elvang} H. Elvang, {\it A charged rotating black ring},
Phys.\ Rev.\ {\bf D68}, 124016 (2003), hep-th/0305247.

\bibitem{pomsen} A.A. Pomeransky and R.A. Sen'kov,
{\it Black ring with two angular momenta}, hep-th/0612005.

\bibitem{hosk} J. Hoskisson,
{\it A charged boubly spinning black ring},
Phys.\ Rev.\ {\bf D79}, 104022 (2009), arXiv:0808.3000 [hep-th].

\bibitem{galsch} D.V. Gal'tsov and N.G. Scherbluk,
{\it Three-charge doubly rotating black ring}, 
Phys.\ Rev.\ {\bf D81}, 044028 (2010), arXiv:0912.2771 [hep-th].

\bibitem{emp2} R. Emparan,
{\it Rotating circular strings, and infinite nonuniqueness of black rings},
JHEP {\bf 0403}, 064 (2004), hep-th/0402149.

\bibitem{emprea2} R. Emparan and H.S. Reall,
{\it Black rings}, 
Class.\ Quant.\ Grav.\  {\bf 23}, R169 (2006), hep-th/0608012.


\end{thebibliography}
\end{document}